\documentclass[preprint,showpacs,preprintnumbers,amsmath,amssymb]{revtex4}

\usepackage{graphicx}% Include figure files
\usepackage{dcolumn}% Align table columns on decimal point
\usepackage{bm}% bold math

\begin{document}

\noindent

\preprint{}

\title{Thermodynamics analogue for self-trapped spinning-stationary Madelung fluid}% Force line breaks with \\

\author{Agung Budiyono}

\affiliation{Institute for the Physical and Chemical Research, RIKEN, 2-1 Hirosawa, Wako-shi, Saitama 351-0198, Japan}

\date{\today}% It is always \today, today,

\begin{abstract} 

We discuss two-dimensional Madelung fluid dynamics whose irrotational case reduces into the Schr\"odinger equation for a free single particle. We show that the self-trapped spinning-stationary Madelung fluid reported in the previous paper can be analogically identified as an equilibrium thermodynamics system. This is done by making correspondence between Shannon entropy over Madelung density and internal energy to be defined in the main text, respectively with thermal-entropy and thermal-internal energy of equilibrium thermodynamics system. This leads us to identify a Madelung fluid analog of thermal-temperature at the vanishing value of which the stationary Madelung fluid will be no more spinning and is equal to the quantum mechanical ground state of a particle trapped inside a cylindrical tube external potential. 

\end{abstract}

%Use showkeys class option if keyword
\maketitle

\section{Introduction: A Class of Self-Trapped Spinning-Stationary Madelung fluid\label{Introduction}}

Let us consider a Madelung fluid dynamics  \cite{Madelung fluid,Takabayashi,Schoenberg} in two-dimensional space, ${\bf q}=\{x,y\}$. The state of the Madelung fluid at time $t$ is determined by a real-valued, scalar, non-negative and normalized function of Madelung density $\rho({\bf q};t)$; and an accompanying velocity vector field ${\bf v}({\bf q};t)$. The two fields are then assumed to satisfy the following coupled nonlinear dynamical equations: 
\begin{eqnarray}
m\frac{d{\bf v}}{dt}=-\partial_{\bf q}U,\hspace{2mm}
\partial_t\rho+\partial_{\bf q}\cdot{\bf j}=0,
\label{Madelung fluid}
\end{eqnarray} 
where $m$ is a parameter of mass dimension. In the above dynamical equation, $U({\bf q};t)$ is generated by $\rho({\bf q};t)$ as
\begin{equation}
U({\bf q};t)=-\frac{\hbar^2}{2m}\frac{\partial_q^2R}{R},
\label{Madelung potential}
\end{equation}
where $R({\bf q};t)=\sqrt{\rho({\bf q};t)}$ and $\partial_q^2=\partial_x^2+\partial_y^2$ is two-dimensional Laplace operator. While ${\bf j}({\bf q};t)$ is the density flow
\begin{equation}
{\bf j}({\bf q};t)=\rho({\bf q};t){\bf v}({\bf q};t). 
\label{density flow}
\end{equation}
Due to its similarity with the Newton equation, the term on the right hand side of the the left equation in (\ref{Madelung fluid}), $-\partial_{\bf q}U$, will be referred to as Madelung force. Accordingly, $U({\bf q};t)$ is called as Madelung potential. 

For particular case of irrotational dynamics where $\partial_{\bf q}\times {\bf v}=0$, one can always write the velocity field as the gradient of some smooth scalar function, $S({\bf q;t})$, as follows:
\begin{equation}
{\bf v}({\bf q};t)=\partial_{\bf q}S/m.
\label{superfluid velocity} 
\end{equation}
Using this to define a complex-valued function $\psi({\bf q};t)=\sqrt{\rho}\exp(iS/\hbar)$, one can then show that Eq. (\ref{Madelung fluid}) can be rewritten into a compact form as:
\begin{equation}
i\hbar\partial_t\psi({\bf q};t)=-\frac{\hbar^2}{2m}\partial_q^2\psi({\bf q};t).
\label{Schroedinger equation}
\end{equation}
This is formally the free Schr\"odinger equation for a single free particle with mass $m$. One however should recall the issue raised by Wallstrom reported in Ref. \cite{Wallstrom objection} on the inequivalence  between the Schr\"odinger equation and the Madelung hydrodynamics equation. He reemphasized that it is the single-valued-ness of the wave function $\psi({\bf q};t)$ in the Schr\"odinger equation which guarantees the old quantization condition. Noticing this and the fact that there is nothing in the context of Madelung fluid which demands a single-valued-ness of the wave function, he then identified that Madelung fluid dynamics can not recover the quantization condition in quantum mechanics. In other words, even for irrotational case, to recover the Schr\"odinger equation from the Madelung fluid one needs to add by hand a quantization condition as in the old quantum theory. 

Let us also note that when the velocity field is irrotational, the pair of equations (\ref{Madelung fluid}) is also the basis for an interpretation of quantum theory called as pilot-wave theory  \cite{Bohm-Hiley book}. In this interpretation, besides the wave function, $\psi$, one assumes the ontological existence of particle trajectory whose dynamics is guided by the wave function through the right equation in Eq. (\ref{Madelung fluid}). Accordingly, $U({\bf q};t)$ given in Eq. (\ref{Madelung potential}) is referred to as quantum potential. Moreover, $\rho({\bf q};t)$ is assumed to give the distribution of the location of the particle. Also, again for irrotational velocity field, adding a nonlinear term of the type $g\rho\psi$, where $g$ is a constant, will give the Gross-Pitaevskii equation describing the condensation of gas of bosons in the mean field regime \cite{Dalfovo review,Cornell-Wieman paper,Ketterle paper}.  

Now let us consider a class of quantum probability densities $\{\rho({\bf q})\}$ having the following form \cite{AgungPRA1,AgungPRA2}:
\begin{equation}
\rho({\bf q})=\frac{1}{Z(T)}\exp\big[-U({\bf q})/T\big], 
\label{canonical Madelung density}  
\end{equation}
where $T$ is a non-negative real number, and $Z(T)$ is a normalization factor given by 
\begin{equation}
Z(T)=\int dq \exp[-U({\bf q})/T], \hspace{2mm}dq=dxdy. 
\label{partition function: position}
\end{equation}
Notice that Eq. (\ref{canonical Madelung density}) together with the definition of Madelung potential given in Eq. (\ref{Madelung potential}) comprise a differential equation for $\rho({\bf q})$ or $U({\bf q})$, subjected to the condition that $\rho({\bf q})$ must be normalized. In term of $U({\bf q})$, one obtains \cite{AgungPRA1,AgungPRA2}
\begin{equation}
\partial_{q}^2U=\frac{1}{2T}(\partial_{\bf q} U\cdot\partial_{\bf q} U)+\frac{4mT}{\hbar^2}U.
\label{NPDE for Madelung potential}
\end{equation}

\begin{figure}[htbp]
\begin{center}
\includegraphics*[width=8cm]{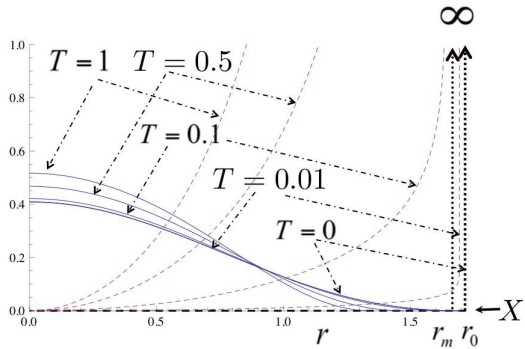}
\end{center}
\caption{The radial profile of rotationally symmetric quantum probability densities (solid lines), each is being self-trapped by its own Madelung potential (dashed lines) satisfying the differential equation (\ref{NPDE for Madelung potential-rotationally symmetric}) for several small values of $T$. The Madelung potential is shifted down so that its minimum equal to zero. We also plot the case for vanishing $T$ given analytically in Eq. (\ref{infinite beta Madelung fluid}). See text for detail.}
\label{rotationally symmetric self-trapped QPD}
\end{figure}

Let us discuss a class of solutions of Eq. (\ref{NPDE for Madelung potential}) which is rotationally symmetric. To do this, it is convenient to use polar coordinate $\{r,\theta\}$, where $r=\sqrt{x^2+y^2}$ and $\theta=\tan^{-1}(y/x)$. Equation (\ref{NPDE for Madelung potential}) can therefore be written as 
\begin{equation}
\partial_r^2U+\frac{\partial_rU}{r}-\frac{(\partial_rU)^2}{2T}-\frac{4mT}{\hbar^2}U=0.
\label{NPDE for Madelung potential-rotationally symmetric}
\end{equation}
Figure \ref{rotationally symmetric self-trapped QPD} shows the numerical solutions of Eq. (\ref{NPDE for Madelung potential-rotationally symmetric}) with the boundary conditions: $U(0)\equiv X=1$ and $\partial_rU(0)=0$, for several small values of $T$. All numerical simulations in this paper is done by setting $\hbar=m=1$. Later we shall vary $X$, yet we will always keep $\partial_qU(0)=0$ fixed in the whole paper. We can see clearly that globally the Madelung density is being trapped by the Madelung potential it itself generates. Moreover, for a given $T$, we found that there is a finite distance $r_m$, at which the Madelung potential is blowing-up \cite{blowing-up NDE}, $U(r_m)=\infty$, so that the Madelung density is vanishing, $\rho(r_m)=0$. Hence, the Madelung density possesses a finite support $\mathcal{M}$ of a disk with radius $r_m(T)$.  

Next, let us write the coupled dynamical equations of Eq. (\ref{Madelung fluid}) in polar coordinate, $\{r,\theta\}$, to give \cite{classical mechanics book}
\begin{eqnarray}
m\Big(\frac{dv_r}{dt}-r\omega^2\Big)=-\partial_rU,\hspace{2mm} \nonumber\\
m\Big(r\frac{d\omega}{dt}+2v_r\omega\Big)=0,\hspace{7mm}\nonumber\\
\partial_t\rho+\partial_r(\rho v_r)=0,\hspace{12mm}
\label{Madelung fluid in polar coordinate}
\end{eqnarray}
where $\omega=d\theta/dt$ is the angular velocity and $v_r=dr/dt$. In the last two lines we have used the fact that our dynamical system is rotationally symmetric. Let us now impose a stationarity condition $v_r=0$. First, from the upmost equation, the angular velocity is related to the Madelung potential as 
\begin{equation}
\omega=\sqrt{\partial_rU/(rm)}.
\label{spinning-stationary condition}
\end{equation}
Since $\partial_rU\neq 0$ for $r\neq 0$, then the Madelung fluid is rotating with radius-dependent angular velocity, $\omega=\omega(r)$. From the middle equation, one gets $d\omega/dt=0$, such that the angular velocity of the Madelung fluid is constant of time. Finally, from the lower most equation, one gets $d\rho/dt=\partial_t\rho=0$. Hence, $\rho(r;t)$ is stationary and spinning around its center with a constant angular velocity field, $\omega(r)$. Physically, the stationarity thus comes from the balance between the attractive Madelung force of the trapping Madelung potential and the repulsive centrifugal force generated by the rotating velocity vector field. Further, note that at the boundary of the support, $r=r_m$, the Madelung potential and Madelung force are infinite. Hence to balance the attractive force, the angular velocity velocity is also infinite. Yet, since the Madelung density is vanishing, one must be assured that the Madelung potential density $U\rho$ and the density flow ${\bf j}$ are also vanishing. 

\section{Equilibrium Thermodynamics Analogy}

In this section, we shall develop analogy between the spinning-stationary Madelung fluid developed in the previous section and equilibrium thermodynamics states. 

\subsection{Internal and kinetic energy}

Next, let us define the energy of the Madelung fluid as follows:
\begin{equation}
{\bar E}\equiv{\bar U}+{\bar K},
\label{total energy}
\end{equation}
where $\bar{U}=\int_{\mathcal{M}} dq\rho({\bf q})U({\bf q})$ is the average Madelung potential and ${\bar K}$ is given by $\bar{K}=\int_{\mathcal{M}} dq\hspace{1mm}\rho({\bf q})K({\bf q})$, where $K({\bf q})=(1/2)m{\bf v}\cdot{\bf v}$. In this respect, ${\bar K}$ should be identified as the {\it kinetic energy} of the Madelung fluid due to the spinning; and accordingly, we shall refer to the average Madelung potential $\bar{U}$ as the {\it  internal energy} of the Madelung fluid, namely the energy possessed by the Madelung fluid when it is not spinning. We showed in Ref. \cite{AgungPRA1} that the kinetic energy can be calculated analytically to give ${\bar K}=T$. One can also show that defining $\psi=\sqrt{\rho}\exp(iS/\hbar)$, the above definition of energy gives equal value as the quantum mechanical average energy over the wave function $\psi$ given by 
\begin{equation}
{\bar E}=\int_{\mathcal{M}} dq\hspace{1mm}\psi^*({\bf q})\Big(-\frac{\hbar^2}{2m}\partial_q^2\Big)\psi({\bf q}). 
\label{quantum mechanical energy}
\end{equation}

\begin{figure}[htbp]
\begin{center}
\includegraphics*[width=8cm]{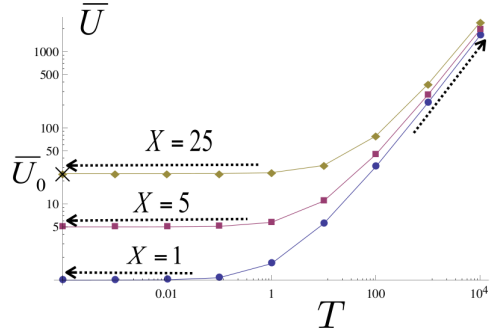}
\end{center}
\caption{The internal energy, $\bar{U}$, of the spinning-stationary Madelung fluid versus $T$ in Log-Log plot for several values of boundary condition $U(0)\equiv X$.}
\label{energy versus beta}
\end{figure}

\begin{figure}[htbp]
\begin{center}
\includegraphics*[width=8cm]{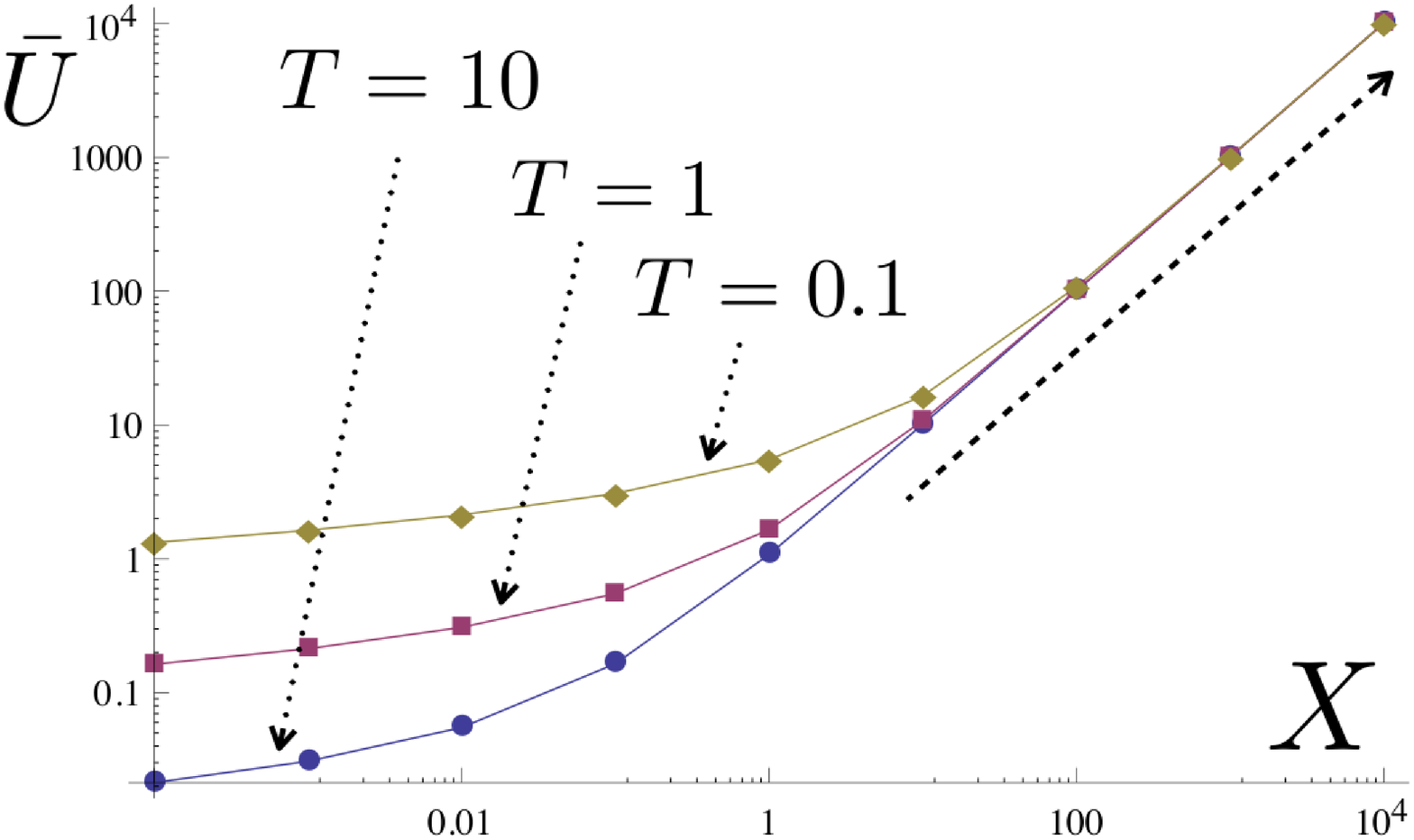}
\end{center}
\caption{The Log-Log plot of the internal energy against the variation of the boundary $U(0)\equiv X$ for several values of $T$.}
\label{u min versus u total}
\end{figure}

Next, let us resort to numerical calculations to discuss the behaviors of the internal energy, $\bar{U}=\bar{U}(T,X)$,  as we vary its parameters $T,X$. Figure \ref{energy versus beta} shows the Log-Log plot of the internal energy $\bar{U}$ against $T$ for the spinning-stationary Madelung fluid which satisfies the differential equation (\ref{NPDE for Madelung potential-rotationally symmetric}) with several values of the boundary condition $X$. One observes that fixing $X$, $\bar{U}$ increases monotonically as we increase $T$. Moreover, for sufficiently large values of $T$, one has the following scaling relation:
\begin{eqnarray}
\bar{U}(T,X)\sim a(X)T,\hspace{3mm}T\gg 1,
\label{scaling U-T}
\end{eqnarray}
where $a$ is positive definite and might depend only on $X$. Hence, for sufficiently large $T$, $\bar{U}$ behaves in similar manner as $\bar{K}$, linearly proportional to $T$. Yet, unlike $\bar{K}$ which depends only on $T$, $\bar{U}$ might depend also on $X$. On the other hand, for vanishing value of $T$, $\bar{U}$ is converging toward a finite value depending on $X$, 
\begin{equation}
\lim_{T\rightarrow 0}\bar{U}\equiv\bar{U}_{0}(X).
\label{vanishing temperature internal energy} 
\end{equation}
Later, we shall show that ${\bar U}_0(X)=X$. 

Figure \ref{u min versus u total} shows the numerical values of the internal energy $\bar{U}$ against $X$ for several values of $T$ in a Log-Log plot. One can see that fixing $T$, the internal energy $\bar{U}$ is monotonically increasing as one increases $X$. One has the following scaling relation for sufficiently small and large values of $X$:
\begin{eqnarray}
\bar{U}(T,X)\sim c(T)X,\hspace{2mm}X\ll 1,\nonumber\\
\bar{U}(T,X)\sim d(T)X^{\beta},\hspace{0mm}X\gg 1,
\label{scaling U-X}
\end{eqnarray}
where $c,d,\beta$ are positive definite. In particular, one can see that for sufficiently large $X$, $\bar{U}$ is independent from $T$. 

$\bar{U}$ thus depends on $T$ and $X$, $\bar{U}=\bar{U}(T,X)$. From Fig. \ref{energy versus beta}, fixing $X$, one can see that $\bar{U}=\bar{U}(T)$ is a one to one mapping. Hence, given a fixed $X$, since $T$ gives ${\bar K}$, then the kinetic energy is completely determined by $\bar{U}$, $\bar{K}=\bar{K}(\bar{U})$. 

\subsection{Madelung fluid analogue for entropy, heat, and temperature}

Let us now develop mathematical relations among nearby (infinitesimally close) spinning-stationary states of the Madelung fluid. To be precise, given a spinning-stationary state with a pair of parameters $(T,X)$ possessing a pair of internal and kinetic energies $(\bar{U},\bar{K})$, we ask: what is the internal and kinetic energies of the infinitesimally close stationary state characterized by a pair of parameters $(T',X')=(T+\delta T,X+\delta X)$. 

To do this, first notice that Eq. (\ref{canonical Madelung density}) can be rewritten as 
\begin{equation}
f\big(U(T,X)\big)dU=\frac{1}{Z}\exp\Big[-\frac{U(T,X)}{T}\Big]\Big|\frac{dr}{dU}\Big|dU. 
\label{the probability of having energy}
\end{equation}
Moreover, the normalization factor of Eq. (\ref{partition function: position}) can then be put as 
\begin{equation}
Z(T,X)=\int_{X}^{\infty} dU\hspace{1mm}\Big|\frac{dr}{dU}\Big|\exp\Big[-\frac{U(T,X)}{T}\Big].
\label{partition function: energy}
\end{equation}
Bearing in mind that $U$ has the dimension of energy, one can see that Eq. (\ref{the probability of having energy}) appears to be in the same fashion as the canonical Maxwell-Boltzmann distribution for equilibrium thermodynamics. 

The above notification leads us to develop an analogy between the class of spinning-stationary Madelung fluid with the setting of equilibrium thermodynamics system giving arise to the canonical ensemble as follows. First, let us consider a thermodynamics system with ``thermal''-internal energy $\bar{U}_s$. Note that $\bar{U}_s$ is a macroscopic quantity assumed to be a statistical average of a microscopic quantity $U_s$ over a probability density $f_s(U_s)$. Next, let us allow the thermal-system to make contact with a thermal-bath of energy $\bar{K}_b$ with finite temperature $T_{b}=T$. The system and the bath can exchange energy such that $\bar{U}_s$ and $\bar{K}_b$ can fluctuate while keeping the total energy conserved. In equilibrium, the temperature of the thermal-system, $T_s$, is then given by the temperature of the bath $T_s=T_{b}=T$ and the probability density that the system possesses the microscopic energy $U_s$ is given by the following canonical ensemble \cite{book on thermodynamics}:
\begin{equation}
f_s(U_s)=\frac{1}{Z_{s}}g(U_s)\exp(-U_s/T),
\label{canonical thermal probability}
\end{equation} 
where $Z_s$ is the normalization constant and $g(U_s)$ is the degree of degeneracy of the energy $U_s$. One thus obtains an analogy between our spinning-stationary state and thermodynamics equilibrium state by making the following correspondence: $\bar{U}\leftrightarrow \bar{U}_s$, $\bar{K}\leftrightarrow \bar{K}_b$ and $|dr/dU|\leftrightarrow g(U_s)$. Namely, one can analogically consider the spinning-stationary Madelung fluid with internal and kinetic energies $(\bar{U},\bar{K})$ developed in the previous section as describing a thermal-system of internal energy $\bar{U}$ in equilibrium with a bath of energy $\bar{K}$ and temperature $T$. Since in our spinning-stationary Madelung fluid state we have $\bar{K}=T$, then the corresponding analog thermal-bath is of ideal gas type. 

\begin{figure}[htbp]
\begin{center}
\includegraphics*[width=7cm]{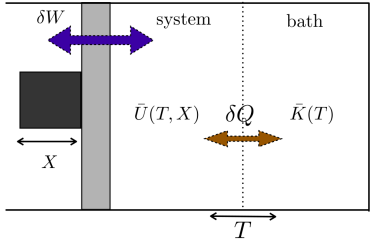}
\end{center}
\caption{System-bath analogy for the spinning-stationary Madelung fluid.}
\label{sys-bath analogy}
\end{figure}

The above system-bath analogy is depicted in Fig. \ref{sys-bath analogy}. To complete the analogy, since the internal energy $\bar{U}$ of the spinning-stationary Madelung fluid depends also on $X$, then one has to connect the analog thermal-system with a piston whose position is determined by $X$. In this way, since the energy of the bath ${\bar K}$ only depends on $T$, then the compression/expansion by the piston can only affect the system through its internal energy. The bath remains unaffected by the variation of $X$. In contrast to this, varying the temperature $T$ will affect both the analog thermal-bath and the system as can be seen pictorially in Fig. \ref{sys-bath analogy}. 

It is thus instructive to apply the equilibrium thermodynamics formalism to our  class of spinning-stationary Madelung fluid to study their parameters dependence. To do this, let us define a new quantity $F$ as follows:
\begin{equation}
-\beta F=\ln Z(\beta,X),
\label{equilibrium thermodynamics 2}
\end{equation} 
where $\beta=1/T$. Taking the difference of the values of $F$ for two infinitesimally close spinning-stationary states, one gets 
\begin{eqnarray}
\delta(-F\beta)=\frac{\partial \ln Z}{\partial\beta}\delta\beta+\frac{\partial
  \ln Z}{\partial X}\delta X\hspace{25mm}\nonumber\\
=-\frac{\delta\beta}{Z}\int_{\mathcal{M}} dq\hspace{1mm}Ue^{-\beta U}-\frac{\beta\delta X}{Z}\int_{\mathcal{M}} dq\hspace{1mm}\frac{\partial U}{\partial X}e^{-\beta U}\hspace{5mm}\nonumber\\ =-\bar{U}\delta\beta-\beta\bar{Y}\delta X=-\delta(\beta\bar{U})+\beta\delta\bar{U}-\beta\bar{Y}\delta X,\hspace{5mm}
\label{equilibrium thermodynamics 3}
\end{eqnarray}
where, in the third equality we have defined a new quantity $Y=\partial U/\partial X$, whose average is given by $\bar{Y}=\int_{\mathcal{M}} dq\hspace{1mm}(\partial U/\partial X)\rho({\bf q})$. Notice that from the differential equation (\ref{NPDE for Madelung potential-rotationally symmetric}), one has $Y>0$, so that ${\bar Y}>0$. Rearranging Eq. (\ref{equilibrium thermodynamics 3}), one finally obtains 
\begin{equation}\
\delta(-F\beta+\bar{U}\beta)=(\delta \bar{U}-\bar{Y}\delta X)\beta.
\label{equilibrium thermodynamics 4}
\end{equation}

Next, let us proceed further to define a new quantity $Q$ such that the difference between its values for two infinitesimally close spinning-stationary states is given by the term inside the bracket on the right hand side of Eq. (\ref{equilibrium thermodynamics 4}) as 
\begin{equation}
\delta Q\equiv\delta \bar{U}-\bar{Y}\delta X.
\label{first law}
\end{equation}
On the other hand, using Eq. (\ref{canonical Madelung density}), one can check that the term inside the bracket on the left hand side of Eq. (\ref{equilibrium thermodynamics 4}) is nothing but equal to the Shannon entropy functional \cite{Shannon entropy} over the Madelung density
\begin{equation}
\beta(-F+\bar{U})=-\int_{\mathcal{M}} dq\hspace{1mm}\rho({\bf q})\ln\rho({\bf q})\equiv H[\rho].
\label{entropy-free energy-quantum energy}
\end{equation}
Later on we shall refer to the above quantity as the Shannon Madelung entropy. In fact, the anzatz of Madelung density given in Eq. (\ref{canonical Madelung density}) can be read as the one that maximizes the above Shannon Madelung entropy $H[\rho]$ given an average Madelung potential or internal energy ${\bar U}$ \cite{Mackey-MEP,Jaynes-MEP}. In this context, it was interpreted in Refs. \cite{AgungPRA1,AgungPRA2} as the most likely Madelung density possessing a finite internal energy. 

Eqs. (\ref{equilibrium thermodynamics 4}), (\ref{first law}) and (\ref{entropy-free energy-quantum energy}) lead us to the following relation between the new quantity $Q$ and Shannon Madelung entropy:
\begin{equation}
\delta Q=\frac{\delta H}{\beta}=T \delta H.
\label{entropy-analog-heat-temperature}
\end{equation}
Finally, inserting Eq. (\ref{entropy-analog-heat-temperature}) back into Eq. (\ref{first law}), one obtains the following relation between the internal energy and Shannon Madelung entropy for two infinitesimally close spinning-stationary states of the Madelung fluid:
\begin{equation}
\delta\bar{U}=T\delta H+\bar{Y}\delta X.
\label{energy-area-entropy}
\end{equation}

\begin{figure}[htbp]
\begin{center}
\includegraphics*[width=7cm]{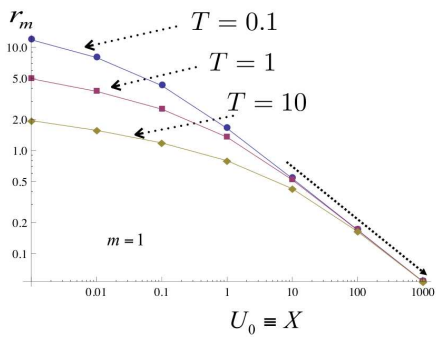}
\end{center}
\caption{The Log-Log plot of the radius of the support $\mathcal{M}$ of the spinning-stationary Madelung fluid, $r_m$, against the values of the boundary $U(0)\equiv X$ for several values of $T$.}
\label{u min versus rmax}
\end{figure}

\begin{figure}[htbp]
\begin{center}
\includegraphics*[width=7cm]{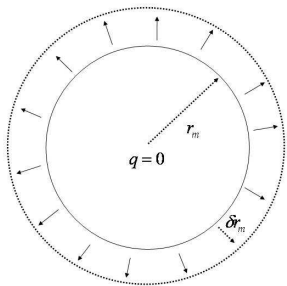}
\end{center}
\caption{Stretching of the boundary.}
\label{bubble stretching}
\end{figure}

A physical interpretation is now  ready  to make. From Eq. (\ref{energy-area-entropy}), since $\delta\bar{U}$ is of energy dimensional, so must be each term on the right hand side. Let us first discuss the physical meaning of the second term, $\delta W \equiv \bar{Y}\delta X$. To do this, let us pay our attention to the behavior of the values of the radius of the support $\mathcal{M}$ of the spinning-stationary Madelung fluid $r_m$, versus the values of $X$. Figure  \ref{u min versus rmax} shows the numerical results of $r_m$ against $X$ for several values of $T$  obtained by solving Eq. (\ref{NPDE for Madelung potential-rotationally symmetric}). One can see clearly that fixing $T$, then $r_m$ is a monotonically decreasing function of $X$. Hence, for a fixed $T$, $r_m(X)$ possesses a unique inverse, $X(r_m)$. One thus obtains a one to one mapping between $X$ and the circumference of the disk support of the spinning-stationary Madelung fluid, $\partial\mathcal{M}$, $L_s(X)=2\pi r_m(X)$. Using this fact, one can rewrite the second term on the right hand side of Eq. (\ref{energy-area-entropy}) as
\begin{eqnarray}
\delta W=\bar{Y}\delta X=-\frac{1}{2\pi}\bar{Y}\Big|\frac{\partial r_m}{\partial X}\Big|_{T}\Big|^{-1}\delta L_s=\sigma_X\delta L_s,\nonumber\\
\mbox{where}\hspace{3mm}\sigma_X\equiv -\frac{1}{2\pi}\bar{Y}\Big|\frac{\partial r_m}{\partial X}\Big|_{T}\Big|^{-1}.\hspace{10mm}
\label{boundary tension}
\end{eqnarray}
Recalling our equilibrium thermodynamics analogy, one can then see the last term on the right hand side of Eq. (\ref{energy-area-entropy}) as the work $\delta W$ needed to stretch the circumference of the boundary of the support of the spinning-stationary Madelung fluid, $\partial\mathcal{M}$, an amount of $L_s$ while keeping $T$ fixed. This situation is depicted in  Fig. \ref{bubble stretching}. 

Next, let us proceed to discuss the first term on the right hand side of Eq. (\ref{energy-area-entropy}), $\delta Q=T\delta H$. It must also be of energy dimensional. Notice that the right hand side is a multiplication of a constant, $T$, with the difference of values of Shannon Madelung entropy for two infinitesimally close spinning-stationary Madelung fluid, $\delta H$. This must remind us to the well-known relation between thermal-heat $Q_{th}$ and thermal-entropy $H_{th}$ in equilibrium thermodynamics, in which one has $\delta Q_{th}=T_{s}\delta H_{th}$. In analogy with this, it is suggestive to call $Q$ and $T$ as the  analog-heat and analog-temperature, respectively. In the context of our equilibrium thermodynamical analogy, $\delta Q$ thus describes the flow of analog-heat from the bath to the system (see Fig. \ref{sys-bath analogy}). 

Finally, in analogy with equilibrium thermodynamics, Eq. (\ref{energy-area-entropy}) can be seen as the {\it law of conversion of energy} in which the increase/decrease of internal energy (average Madelung potential) $\delta \bar{U}$ is equal to the amount of work done as the support of the spinning-stationary Madelung fluid expanding/shrinking, $\delta W$, plus the amount of incoming/outgoing analog-heat, $\delta Q$. 

\begin{figure}[htbp]
\begin{center}
\includegraphics*[width=7cm]{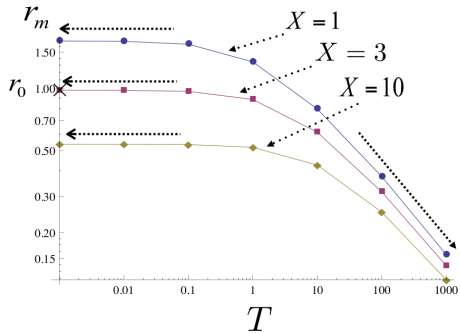}
\end{center}
\caption{$r_m$ against $T$ for several values of $X$.}
\label{radius of support}
\end{figure}

Next, from Eq. (\ref{entropy-free energy-quantum energy}), $F$ defined in Eq. (\ref{equilibrium thermodynamics 2}) must be identified as the Madelung fluid analog of thermodynamics free energy. It is given by
\begin{equation}
F={\bar U}-TH. 
\label{free energy}
\end{equation}
The difference of its values for two infinitesimally close spinning-stationary states is given by\begin{equation}
\delta F=\sigma_X\delta L_s-H\delta T,
\label{variation of free energy}
\end{equation}
where we have used Eqs. (\ref{energy-area-entropy}) and (\ref{boundary tension}). Hence it is a function of the analog-temperature $T$ and the circumference of the support of the spinning-stationary Madelung fluid, $L_s(X)$. On the other hand, Fig. \ref{radius of support} shows a numerical evidence that $L_s=2\pi r_m$ depends also on $T$, $L_s=L_s(T,X)$. In particular, fixing $X$, $L_s$ is a monotonic decreasing function of $T$. One can then rewrite Eq. (\ref{variation of free energy}) as 
\begin{eqnarray}
\delta F=(\sigma_X-\sigma_{T})\delta L_s\equiv\sigma\delta L_s,\hspace{5mm}\nonumber\\
\mbox{where}\hspace{3mm}\sigma_{T}=H\frac{\partial T_s}{\partial L_s}\Big|_{X},\hspace{2mm} \sigma\equiv\sigma_X-\sigma_{T}.
\label{free energy - circumference}
\end{eqnarray}
Hence, $\delta F$ can be interpreted as the total work needed to expand the support of the stationary-spinning  Madelung fluid by varying $T$ and/or $X$. In this sense, $\sigma$ can be regarded as the tension along the boundary of the support of the spinning-stationary Madelung fluid, $\partial\mathcal{M}$. Moreover, since, the class of Madelung density of Eq. (\ref{canonical Madelung density}) is the one that maximizes $H[\rho]$ given ${\bar U}$, then from Eq. (\ref{free energy}), it is also the one that minimizes $F$. 

Proceeding in this way, one can further derive various relations valid for our spinning-stationary Madelung fluid, in analogy with the various relations familiar in equilibrium thermodynamics.  

\subsection{Quantum mechanical ground state as the limit of vanishing analog-temperature}

Let us pay our attention again on Fig. \ref{radius of support}. Two limiting cases are of great interest. First, one sees that at infinite value of $T$, $r_m$ is vanishing, $\lim_{T\rightarrow \infty}r_m=0$. Hence, one can conclude that in the limit $T\rightarrow\infty$, the Madelung density is approaching a delta function with singular support \cite{AgungPRA1}
\begin{equation}
\lim_{T\rightarrow \infty}\rho({\bf q};X,T)=\delta({\bf q}), 
\end{equation}
regardless of the value of $X$. Moreover one has $\lim_{T\rightarrow\infty}\bar{K}=\lim_{T\rightarrow\infty}T\rightarrow \infty$. Numerical results depicted in Fig. \ref{energy versus beta} also shows that in the limit of infinite analog-temperature $T$, the internal energy is approaching infinity, $\lim_{T\rightarrow\infty}\bar{U}\rightarrow\infty$. Hence, at infinite analog-temperature the total energy is also infinite $\lim_{T\rightarrow\infty}{\bar E}=\infty$. For this reason, this limit is physically irrelevant.  

Next, let us discuss the other limiting case in which the analog-temperature is vanishing, $T\rightarrow 0$.  Numerical results in Fig. \ref{radius of support} shows that the radius of the support of the spinning-stationary Madelung fluid is converging toward certain finite value depending on the value of $X$,
\begin{equation}
\lim_{T\rightarrow 0}r_m(T,X)= r_0(X).
\label{radius of support at infinite beta}
\end{equation}
This shows that the Madelung density and its corresponding Madelung potential are converging toward certain functions
\begin{equation}
\lim_{T\rightarrow 0}\rho({\bf q};T,X)=\rho_{0}({\bf q};X), \hspace{2mm}\lim_{T\rightarrow 0}U({\bf q};T,X)=U_0({\bf q};X).
\label{infinite beta Madelung density}
\end{equation} 
In fact, one can see in Fig. \ref{rotationally symmetric self-trapped QPD} that as $T$ decreases, the Madelung potential is getting flatterer inside the support before becoming infinite at the boundary line $r=r_m$. Hence, one can expect that as $T$ is approaching toward zero, then $U(r)$ is approaching  a cylinder of perfectly flat bottom with infinite (hard) wall at its boundary $r=r_0$. 

Let us  show that our expectation above is correct. To do this, at $T=0$, let us assume that the Madelung potential inside the support is constant given by $U_0(0)=X$ (see Fig. \ref{rotationally symmetric self-trapped QPD}) and infinite at the boundary line $U_0(r_0)=\infty$. Since the kinetic energy is vanishing, one has 
\begin{eqnarray}
{\bar E}=\bar{U}_{0}=\int_{\mathcal{M}} dq\hspace{1mm}\rho_0({\bf q})X=X.\end{eqnarray}
Hence, recalling the definition of Madelung potential given in Eq. (\ref{Madelung potential}), inside $\mathcal{M}$, one has
\begin{equation}
-\frac{\hbar^2}{2m}\partial_q^2R_0({\bf q};{\bar E})=XR_0({\bf q};{\bar E})={\bar E} R_0({\bf q};{\bar E}), 
\label{infinite beta Madelung fluid 1}
\end{equation}
where we have denoted the quantum amplitude at vanishing $T$ by $R_0({\bf q};{\bar E})\equiv\sqrt{\rho_0({\bf q};{\bar E})}$. The above differential equation must be subjected to the boundary condition that along the boundary line of the support, $\partial\mathcal{M}$, the Madelung potential is infinite so that the Madelung density is vanishing: $\rho_0(r_0;{\bar E})=0$. 

Equation (\ref{infinite beta Madelung fluid 1}) with the boundary condition described above is nothing but the stationary (time-independent) Schr\"odinger equation for a particle of mass $m$ trapped inside a cylindrical tube external potential whose bottom is flat and boundary is infinitely high. In the latter case, $R_0({\bf q};{\bar E})$ is called as the single particle quantum mechanical ground state. In polar coordinate, recalling that the Madelung fluid is rotationally symmetric, one obtains the following differential equation
\begin{equation}
\partial_r^2R_0(r;{\bar E})+\frac{1}{r}\partial_rR_0(r;{\bar E})=-\frac{2m{\bar E}}{\hbar^2}R_0(r;{\bar E}),
\label{infinite beta Madelung fluid 3}
\end{equation}
the solution of which is given by 
\begin{equation}
R_0(r;{\bar E})=A{\mathcal J}(0,kr),
\label{infinite beta Madelung fluid}
\end{equation}
where $\mathcal{J}(0,r)$ is Bessel function of the first kind, $k=\sqrt{2m{\bar E}/\hbar^2}$ and $A$ is a normalization constant. The boundary condition implies the relation between the energy and the radius of the support as $kr_0=B_0$, where $B_0$ is the first zero of the Bessel function. 

Figure \ref{rotationally symmetric self-trapped QPD} indeed confirms that as $T$ is decreasing toward zero, $\rho(r;T,{\bar E})$ obtained by solving the differential equation (\ref{NPDE for Madelung potential-rotationally symmetric}) is converging toward $\rho_0(r;{\bar E})$ given analytically in Eq. (\ref{infinite beta Madelung fluid}). This suggests that our initial guess that the Madelung potential at the vanishing spinning temperature takes the form of cylinder tube with flat bottom and infinite hard wall at the boundary is correct. Moreover, since ${\bar K}=T$, then at the limit of vanishing analog-temperature $T=0$, the kinetic energy of the Madelung fluid is vanishing, ${\bar K}=0$. In other words, the  Madelung fluid is no more spinning so that the repulsive centrifugal force is also vanishing. On the other hand,   since the Madelung potential is perfectly flat inside the support, then the Madelung force, $-\partial_{\bf q}U$, is also vanishing. This ensures that $\rho_0(r)$ is a non-spinning yet stationary solution of the two dimensional Madelung fluid. Further, since the Madelung fluid is static, Eq. (\ref{Madelung fluid}) is now equal to the Schr\"odinger equation of (\ref{Schroedinger equation}). In this case, the complex-valued wave function takes the form $\psi_0({\bf q};{\bar E})\equiv R_0({\bf q};{\bar E})\exp(iS_0/\hbar)$, where $S_0$ is arbitrary constant. 

\section{Conclusion and discussion}

First, we have discussed a class of self-trapped, spinning-stationary solutions of two-dimensional Madelung fluid  reported in the previous paper \cite{AgungPRA1}. In particular, the energy of the Madelung fluid is defined as the summation of the average Madelung potential ${\bar U}$ and average kinetic energy ${\bar K}$. We then further identify the average Madelung potential as the internal energy of the system, namely the energy when the fluid is not spinning. Surprisingly, the profile of the Madelung density of the spinning-stationary Madelung fluid turns out to be in similar form as Maxwell-Boltzmann canonical distribution in equilibrium thermodynamics, namely the Madelung density is given as the negative exponentiation of Madelung potential (whose average gives the internal energy), divided by some non-negative parameter. 

The above facts enables us to develop analogical correspondence between the spinning-stationary Madelung fluid  with the setting of equilibrium thermodynamics giving arise to the canonical ensemble. This is done by considering a thermal system of internal energy $\bar{U}$ in equilibrium with a thermal bath of energy $\bar{K}$ with finite temperature, $T$. In this analogy, the Shannon Madelung entropy  corresponds to thermal-entropy of the thermal system. Moreover, since ${\bar K}=T$, we refer to $T$ as the analog-temperature. This also allows us to define a new quantity called as analog-heat which is given by the multiplication between the analog-temperature and Shannon Madelung entropy.    

By comparing two infinitesimally close spinning-stationary states of the Madelung fluid, we proceeded to derive mathematical relations in phenomenological analogy with relations familiar in equilibrium thermodynamics. The analogy includes Madelung fluid version of thermodynamics first law on the conversion of energy. In our spinning-stationary Madelung fluid, we showed that the increase/decrease of the internal energy of the spinning-stationary Madelung fluid is equal to the work done to expand/compress the support of spinning-stationary Madelung fluid plus the gain/loss of the analog-heat. We also showed that at the vanishing analog-temperature $T$, the square root of the Madelung density is equal to  the quantum mechanical ground state of a single particle trapped in a cylindrical tube external potential.   

Two immediate interesting question are thus in order. First, since allowing singularity in quantum phase defined in Eq. (\ref{superfluid velocity}) will reduce the Madelung fluid dynamics of Eq. (\ref{Madelung fluid}) into the Schr\"odinger equation of (\ref{Schroedinger equation}), then it is interesting to ask if there is an anomaly quantum system with singular phase whose state is given by the spinning-stationary state discussed in this paper. In this case, one is interested to know the quantum mechanical meaning of the analog-temperature, $T$. In particular, in this quantum system, the Shannon Madelung entropy becomes the ordinary  Shannon information entropy. Second, since in the vanishing of analog-temperature, the stationary Madelung state satisfies the time-independent Schr\"odinger equation, then the stationary non-spinning Madelung fluid discussed in the previous section must be realizable in quantum system worthy experimental observation. In this sense, it is then interesting to ask if a quantum state can self-trap itself with self-generated boundary even in the absence of external potential. 

\begin{acknowledgments}

This research is partially funded by FPR program in RIKEN. The author acknowledges useful discussion with Ken Umeno. 

\end{acknowledgments}

\end{document}